\documentclass[12pt]{article}

\usepackage{amsmath}
\usepackage{amssymb}
\usepackage{amsfonts}
\usepackage{latexsym}
\usepackage{color}
\usepackage[english]{babel}

\usepackage{graphicx}
\usepackage{subfigure}

\usepackage{epstopdf}
\catcode `\@=11 \@addtoreset{equation}{section}

\catcode `\@=12

\newcommand{\be}{\begin{equation}}
	\newcommand{\en}{\end{equation}}
\newcommand{\bea}{\begin{eqnarray}}
	\newcommand{\ena}{\end{eqnarray}}
\newcommand{\beano}{\begin{eqnarray*}}
	\newcommand{\enano}{\end{eqnarray*}}
\newcommand{\bee}{\begin{enumerate}}
	\newcommand{\ene}{\end{enumerate}}

\newcommand{\mc}{\mathcal}

\newcommand{\D}{{\mc D}}

\newcommand{\Sc}{{\cal S}}

\newcommand{\F}{{\cal F}}
\newcommand{\G}{{\cal G}}

\newcommand{\Lc}{{\cal L}}
\newcommand{\ltwo}{{\Lc^2(\mathbb{R})}}
\newcommand{\scr}{{\Sc(\mathbb{R})}}

\newcommand{\1}{1 \!\! 1}

\newcommand{\Hil}{\mc H}

\newtheorem{thm}{Theorem}

\newtheorem{prop}[thm]{Proposition}
\newtheorem{defn}[thm]{Definition}

\catcode `\@=11 \@addtoreset{equation}{section}
\catcode `\@=12

\begin{document}

\thispagestyle{empty}

\vspace*{2cm}

\begin{center}
{\Large \bf A fully pseudo-bosonic Swanson model}   \vspace{2cm}\\

{\large Fabio Bagarello}\\
  Dipartimento di  Ingegneria, \\
Universit\`a di Palermo,\\ I-90128  Palermo, Italy, and\\
INFN, Catania\\
e-mail: fabio.bagarello@unipa.it\\

\end{center}

\vspace*{1cm}

\begin{abstract}

We consider a fully pseudo-bosonic Swanson model  and we show how its Hamiltonian $H$ can be diagonalized. We  also deduce the eigensystem of $H^\dagger$, using the general framework and results deduced in the context of pseudo-bosons. We also construct, using different approaches, the bi-coherent states for the model, study some of their properties, and compare the various constructions.

\end{abstract}

\vspace{2cm}

%{\bf PACS Numbers}:  .......

\vfill

%\pagenumbering{roman}

\newpage

\section{Introducion}\label{sectintro}

In the standard literature on quantum mechanics one of the main axioms of any {\em well established} approach to the analysis of the microscopic world is that the observables of a physical system, $\Sc$, are represented by self-adjoint operators. This is, in particular, what is required to the Hamiltonian of $\Sc$. Since few decades, however, it has become more and more evident that this is only a sufficient condition to require, but it is not also necessary. This (apparently) simple remark is at the basis of thousands of papers, and of many monographs. Here we only cite some of these latter, \cite{specissue2012}-\cite{specissue2021}, where many other references can be found. 

The use of non self-adjoint Hamiltonians opens several lines of research, both for its possible implications in Physics, and for the mathematical issues raised by this extension. In particular, in a series of papers and in the books \cite{baginbagbook,bagspringer}, a particular class of non self-adjoint Hamiltonians has been analyzed in detail, together with their connections with a special class of coherent states. These Hamiltonians are constructed in terms of {\em pseudo-bosonic operators} which are,  essentially, suitable deformations of the bosonic creation and annihilation operators. These deformations are again ladder operators, and this is the reason why we were, and still are, interested in finding the eigenstates of these new annihilation operators. Several examples have been constructed along the years, by us and by other authors, {\cite{tri}-\cite{dapro}}. In particular, one Hamiltonian which has become very famous in the literature on $PT$-quantum mechanics is related to what is now called {\em the Swanson model}, \cite{dapro,swan,bagswa}. The Hamiltonian for this model is $H_s=\omega_s c^\dagger c+\alpha c^2+\beta {c^\dagger}^2$, where $\alpha, \beta, \omega_s\in\mathbb{R}$ and where $[c,c^\dagger]=\1$. Of course, since $c$ and $c^\dagger$ are unbounded, the above expressions for $H_s$ and  $[c,c^\dagger]$ are simply formal. To make them rigorous, we should add, in particular, details on their domains of definition. A more mathematical approach to $H_s$, closer to what is relevant for us here, can be found in \cite{bagswa,baginbagbook,bagspringer}. In particular, we have shown that $H_s$ can be rewritten in a diagonal form in terms of pseudo-bosonic operators, and this has been used to analyze in detail its spectrum and its eigenvectors. In particular, we have shown that the set of these eigenvectors is complete in $\Hil$, but it is not a basis. There are many papers devoted to the Swanson model, in one of its various expressions. Some other paper on this model are the following: \cite{sinharoy}-\cite{baginvertedosc}, just to cite a few.

In this paper we will focus on a particular version of a fully pseudo-bosonic extension of $H_s$, i.e. on a version in which the pair of bosonic operators $(c,c^\dagger)$ are replaced, from the very beginning,  by operators $(a,b)$ satisfying certain properties, see Section \ref{sectprel}. Moreover, to simplify the general treatment, and without any particular loss of generality, we will also restrict to choosing $\alpha=\beta$. Notice that, while this choice trivializes the original model, in the sense that $H_s=H_s^\dagger$, does not change at all the lack of self-adjointness of the Hamiltonian $H$ we will introduce later, see (\ref{31}).

The paper is organized as follows: after a review on pseudo-bosons, in Section \ref{sectprel}, we propose our {\em fully pseudo-bosonic Swanson model}, and we find the eigenvalues and the eigenvectors of the Hamiltonian of the system, and of its adjoint. We prove that the sets of these eigenvectors are complete and biorthonormal in $\ltwo$, while they are not bases. This will be done in Section \ref{sectmodel}. Section \ref{sectbcs} is focused on bi-coherent states and on their properties. Section \ref{sectconcl} contains our conclusions, and plans for the future.

\section{Preliminaries}\label{sectprel}

This section is devoted to some preliminary definitions and results on pseudo-bosons (PBs). This will be needed in the next sections, where the modified Swanson Hamiltonianl will be introduced and analyzed.

 Let $a$ and $b$ be two operators
on $\Hil$, with domains $\D(a)$ and $\D(b)$ respectively, $a^\dagger$ and $b^\dagger$ their adjoint, and let $\D$ be a dense subspace of $\Hil$, stable under the action of $a$, $b$ and their adjoints. It is clear that $\D\subseteq \D(a^\sharp)$ and $\D\subseteq \D(b^\sharp)$, where $c^\sharp=c,c^\dagger$, and that $a^\sharp f,b^\sharp f\in\D$ for all $f\in\D$. Then both $abf$ and $baf$ are well defined, $\forall f\in\D$.

\begin{defn}\label{def21}
	The operators $(a,b)$ are $\D$-pseudo-bosonic ($\D$-pb) if, for all $f\in\D$, we have \be a\,b\,f-b\,a\,f=f. \label{21}\en
\end{defn}
Sometimes, to simplify the notation, rather than (\ref{21}) one writes $[a,b]=\1$. It is not surprising that neither $a$ nor $b$ are bounded on $\Hil$. This is the reason why the role of $\D$ is so relevant, here and in the rest of these notes.

\vspace{2mm}

Our working assumptions for dealing with these operators are the following:

\vspace{2mm}

{\bf Assumption $\D$-pb 1.--}  there exists a non-zero $\varphi_{ 0}\in\D$ such that $a\,\varphi_{ 0}=0$.

\vspace{1mm}

{\bf Assumption $\D$-pb 2.--}  there exists a non-zero $\Psi_{ 0}\in\D$ such that $b^\dagger\,\Psi_{ 0}=0$.

\vspace{2mm}

Notice that, if $b=a^\dagger$, then these two assumptions collapse into a single one and (\ref{21}) becomes the well known canonical commutation relations (CCR), for which the existence of a vacuum which belongs to an invariant set ($\scr$, for instance) is guaranteed. Then, for CCR, Assumptions $\D$-pb 1 and $\D$-pb 2 are automatically true.

In \cite{bagspringer} it is widely discussed the possibility that $[a,b]=\1$ can be extended outside $\ltwo$. This gives rise, as we will briefly comment later in Section \ref{sect21}, to the so-called weak PBs (WPBs), in which a central role is no longer played by $\ltwo$, but by other functional spaces. 

\vspace{2mm}

In the present situation, the stability of $\D$ under the action of $b$ and $a^\dagger$ implies, in particular, that $\varphi_0\in
D^\infty(b):=\cap_{k\geq0}D(b^k)$ and that $\Psi_0\in D^\infty(a^\dagger)$. Here $D^\infty(X)$ is the domain of all the powers of the operator $X$. Hence \be
\varphi_n:=\frac{1}{\sqrt{n!}}\,b^n\varphi_0,\qquad \Psi_n:=\frac{1}{\sqrt{n!}}\,{a^\dagger}^n\Psi_0, \label{FB32}\en $n\geq0$, are well defined
vectors in $\D$ and, therefore, they belong to the domains of $a^\sharp$, $b^\sharp$ and $N^\sharp$, where $N=ba$ and $N^\dagger$ is the adjoint of $N$. We introduce next the sets $\F_\Psi=\{\Psi_{ n}, \,n\geq0\}$ and $\F_\varphi=\{\varphi_{ n}, \,n\geq0\}$.

It is now simple to deduce the following \index{Relation!lowering}lowering and \index{Relation!raising}raising relations:
\be
\left\{
\begin{array}{ll}
	b\,\varphi_n=\sqrt{n+1}\,\varphi_{n+1}, \qquad\qquad\quad\,\, n\geq 0,\\
	a\,\varphi_0=0,\quad a\varphi_n=\sqrt{n}\,\varphi_{n-1}, \qquad\,\, n\geq 1,\\
	a^\dagger\Psi_n=\sqrt{n+1}\,\Psi_{n+1}, \qquad\qquad\quad\, n\geq 0,\\
	b^\dagger\Psi_0=0,\quad b^\dagger\Psi_n=\sqrt{n}\,\Psi_{n-1}, \qquad n\geq 1,\\
\end{array}
\right.
\label{FB33}\en as well as the  eigenvalue equations $N\varphi_n=n\varphi_n$ and   $N^\dagger\Psi_n=n\Psi_n$, $n\geq0$, where, more explicitly,
$N^\dagger=a^\dagger b^\dagger$.  Incidentally we observe that this last equality should be understood, here and in the following, on $\D$: $N^\dagger
f=a^\dagger b^\dagger f$, $\forall\, f\in\D$.

As a consequence
of these  equations,  choosing the normalization of $\varphi_0$ and $\Psi_0$ in such a way $\left<\varphi_0,\Psi_0\right>=1$, it is easy to show that
\be \left<\varphi_n,\Psi_m\right>=\delta_{n,m}, \label{FB34}\en
for all $n, m\geq0$. The conclusion is, therefore, that $\F_\varphi$ and $\F_\Psi$
are biorthonormal sets of eigenstates of $N$ and $N^\dagger$, respectively. Notice that these latter operators, which are manifestly non self-adjoint if $b\neq a^\dagger$, have both non negative integer eigenvalues and, because of that, they are called {\em number-like} (or simply number) operators.   The properties we have deduced for $\F_\varphi$ and $\F_\Psi$, in principle, does not allow us to conclude that they are (Riesz) bases or not for $\Hil$. In fact, this is not always
the case, \cite{bagspringer}, even if sometimes (for {\em regular} PBs, see below), this is exactly what happens. With this in mind, let us introduce for the following assumption:

\vspace{2mm}

{\bf Assumption $\D$-pb 3.--}  $\F_\varphi$ is a basis for $\Hil$.

\vspace{1mm}

\noindent This is equivalent to assume that $\F_\Psi$ is a basis as well, \cite{chri,you}.   While Assumption $\D$-pb 3, is not always satisfied, in most of the concrete situations considered so far in the literature, it is true that $\F_\varphi$ and $\F_\Psi$ are total in $\Hil=\ltwo$. For this reason, it is more reasonable to replace Assumption $\D$-pb 3 with this weaker version:

\vspace{2mm}

{\bf Assumption $\D$-pbw 3.--}  $\F_\varphi$ and $\F_\Psi$ are $\G$-quasi bases, for some subspace $\G$ dense\footnote{Notice that $\G$ does not need to coincide with $\D$, even if sometimes this happens.} in $\Hil$.

\vspace{2mm}

This means that, $\forall f,g\in\G$, the following identities hold
\be
\left<f,g\right>=\sum_{n\geq0}\left<f,\varphi_n\right>\left<\Psi_n,g\right>=\sum_{n\geq0}\left<f,\Psi_n\right>\left<\varphi_n,g\right>. \label{FB35}
\en

It is obvious that, while Assumption $\D$-pb 3 implies (\ref{FB35}), the reverse is false. However, if $\F_\varphi$ and $\F_\Psi$ satisfy
(\ref{FB35}), we still have some (weak) form of \index{Identity!resolution of the}resolution of the identity, and, from a physical and from a mathematical point of view, this is enough to deduce
interesting results. For instance, if $f\in \G$ is orthogonal to all the $\Psi_n$'s (or to all the $\varphi_n$'s), then $f$
is necessarily zero: $\F_\Psi$ and $\F_\varphi$ are total in $\G$. Indeed, using (\ref{FB35}) with $g=f\in\G$, we find $\|f\|^2=\sum_{n\geq0}\left<f,\varphi_n\right>\left<\Psi_n,f\right>=0$
since $\left<\Psi_n,f\right>=0$ (or $\left<f,\varphi_n\right>=0$) for all $n$. But, since $\|f\|=0$, then $f=0$.

For completeness we briefly discuss the role of two  \index{Operator!intertwining}intertwining operators which are intrinsically related to our $\D$-PBs. We only consider the regular case here. More details can be found in \cite{baginbagbook}.

In the regular case, Assumption $\D$-pb 3 holds in a strong  form: $\F_\varphi$ and $\F_\Psi$ are biorthonormal Riesz bases, so that we have
\be
f=\sum_{n=0}^\infty\left<\varphi_n,f\right>\,\Psi_n=\sum_{n=0}^\infty\left<\Psi_n,f\right>\,\varphi_n,
\label{FB36}\en
$\forall f\in\Hil$. Looking at these expansions, it is natural to ask if sums like $S_\varphi f=\sum_{n=0}^\infty\left<\varphi_n,f\right>\,\varphi_n$ or $S_\Psi f=\sum_{n=0}^\infty\left<\Psi_n,f\right>\,\Psi_n$ also make some sense, or for which vectors  they do converge, if any. In our case, since $\F_\varphi$ and $\F_\Psi$ are Riesz bases, we know that an orthonormal basis $\F_e=\{e_n\}$ exists, together with a bounded operator $R$ with bounded inverse, such that $\varphi_n=Re_n$ and $\Psi_n=(R^{-1})^\dagger e_n$, $\forall n$. It is clear that, if $R=\1$, all these sums collapse and converge to $f$. But, what if $R\neq\1$?

The first result follows from the biorthonormality of $\F_\varphi$ and $\F_\Psi$, which easily implies that
\be S_\varphi\Psi_{ n}=\varphi_{ n},\qquad S_\psi\varphi_{ n}=\Psi_{ n},
\label{FB37}\en for all $ n\geq0$. These equalities, which are true for biorthogonal bases non necessarily of the Riesz type,
together imply that $\Psi_{ n}=(S_\Psi\,S_\varphi)\Psi_{ n}$ and $\varphi_{ n}=(S_\varphi
\,S_\Psi)\varphi_{ n}$, for all $ n\geq0$. These formulas, in principle, cannot be extended to all of $\Hil$ except when $S_\varphi$ and $S_\Psi$
are bounded. If this is the case, then we deduce that
\be S_\Psi\,S_\varphi=S_\varphi\,S_\Psi=\1 \quad \Rightarrow \quad S_\Psi=S_\varphi^{-1}. \label{FB38}\en In other words, both $S_\Psi$
and $S_\varphi$ are invertible and one is the inverse of the other. This is what happens, in particular, for regular $\D$-PBs. In fact, in this situation, it is possible to relate $S_\varphi$ and $S_\psi$ with the operator $R$ connecting $\F_e$ with $\F_\varphi$ and $\F_\Psi$:  let $f\in D(S_\varphi)$, which for the moment we do not assume to be coincident with $\Hil$. Then
$$
S_\varphi f:=\sum_{ n} \left<\varphi_{n},f\right>\,\varphi_{ n}=\sum_{ n} \left<Re_{
	n},f\right>\,Re_{ n}=R\left(\sum_{ n} \left<e_{
	n},R^\dagger f\right>\,e_{ n}\right)=RR^\dagger f,
$$
where we have used the facts that $\F_e$ is an orthonormal basis and that $R$ is bounded and, therefore, continuous. Of course
$RR^\dagger$ is bounded as well and the above equality can be extended to all of $\Hil$. Therefore we conclude that $S_\varphi=RR^\dagger$. In a similar way we
can deduce that $S_\Psi=(R^\dagger)^{-1}R^{-1}=S_\varphi^{-1}$, which is also bounded. In fact, using the C*-property for $B(\Hil)$, we deduce
that $\|S_\varphi\|=\|R\|^2$ and $\|S_\Psi\|=\|R^{-1}\|^2$. It is also clear that $S_\varphi$ and $S_\Psi$ are positive operators, and it is interesting to check that they satisfy the following intertwining relations:
\be
S_\Psi N\varphi_n=N^\dagger S_\Psi\varphi_n, \qquad N S_\varphi\Psi_n=S_\varphi N^\dagger\Psi_n,
\label{FB39}\en
Indeed we have, recalling that $N\varphi_n=n\varphi_n$ and $N^\dagger\Psi_n=n\Psi_n$, $S_\Psi N\varphi_n=n(S_\Psi\varphi_n)=n\Psi_n$, as well as $N^\dagger S_\Psi\varphi_n=N^\dagger\Psi_n=n\Psi_n$. The second equality in (\ref{FB39}) follows from the first one, simply by left-multiplying  $S_\Psi N\varphi_n=N^\dagger S_\Psi\varphi_n$ with $S_\varphi$, and using (\ref{FB37}). These relations are not surprising, since intertwining relations can be often established between operators having the same eigenvalues.

The situation is mathematically much more complicated for $\D$-PBs which are not regular. This is mainly because there is no reason for $S_\varphi$ and $S_\Psi$ to be bounded, or for the series  $\sum_{n=0}^\infty\left<\varphi_n,f\right>\,\varphi_n$ and $\sum_{n=0}^\infty\left<\Psi_n,f\right>\,\Psi_n$ (which are those used to { define} these operators) to be convergent, at least on some dense set. We refer to \cite{baginbagbook,bagspringer} for more results on this and other aspects of PBs. It is also useful to stress that these operators are connected to what, mostly in the physical literature, are called the {\em metric operators}, often appearing in connection with $PT$-symmetric Hamiltonians, \cite{benbook,metr2,metr3}.

\subsection{Leaving $\ltwo$}\label{sect21}

Also in view of what will be discussed later in this paper, we are  interested now in considering first order differential operators of the form
\be
a=\alpha_a(x)\,\frac{d}{dx}+\beta_a(x), \qquad b=-\frac{d}{dx}\,\alpha_b(x)+\beta_b(x), 
\label{41}\en
for some suitable $C^\infty$ functions $\alpha_j(x)$ and $\beta_j(x)$, $j=a,b$, \cite{bagJMAA}, where we have shown that 
 these operators produce, using the strategy outlined before, two families of functions, $\F_\varphi=\{\varphi_{n}(x)\}$ and $\F_\Psi=\{\Psi_{n}(x)\}$, which may, or may not, be square-integrable. More results on this particular class of PBs are also given in 
 \cite{bagJPA2020,bagJPCS2021,bagspringer}.

We first compute $[a,b]$ on some sufficiently regular function $f(x)$, not necessarily in $\ltwo$. For what we need, it is sufficient to assume $f(x)$ to be at least $C^2$. Of course, this requirement could be relaxed if we interpret $\frac{d}{dx}$ as the weak derivative, but this will not be done here.
An easy computation shows that, under this mild condition on $f(x)$, $[a,b]f(x)$ does make sense, and $[a,b]f(x)=f(x)$ if $\alpha_j(x)$ and $\beta_j(x)$, $j=a,b$, satisfy the equalities
\be
\left\{
\begin{array}{ll}
	\alpha_a(x)\alpha_b'(x)=\alpha_a'(x)\alpha_b(x), \\
	\alpha_a(x)\beta_b'(x)+	\alpha_b(x)\beta_a'(x)=1+\alpha_a(x)\alpha_b''(x).\\
\end{array}
\right.
\label{42}\en
In particular, the first equality is always true if $\alpha_a(x)$ and $\alpha_b(x)$ are both constant, as it will be the case for our model, see (\ref{38}). In general,  it is convenient to assume that they are never zero: $\alpha_j(x)\neq0$, $\forall x\in\mathbb{R}$, $j=a,b$. 

Under this assumption, it is easy to find the vacua of $a$ and of $b^\dagger$, as required in Assumptions $\D$-pb1 and $\D$-pb2. Here 
\be
a^\dagger=-\frac{d}{dx}\,\overline{\alpha_a(x)}+\overline{\beta_a(x)}, \qquad b^\dagger=\overline{\alpha_b(x)}\,\frac{d}{dx}+\overline{\beta_b(x)}.
\label{43}\en

The vacua of $a$ and $b^\dagger$ are the solutions of the equations $a\varphi_0(x)=0$ and $b^\dagger\psi_0(x)=0$, which are easily found:
\be
\varphi_0(x)=N_\varphi \exp\left\{-\int\frac{\beta_a(x)}{\alpha_a(x)}\,dx\right\}, \qquad \psi_0(x)=N_\psi \exp\left\{-\int\frac{\overline{\beta_b(x)}}{\overline{\alpha_b(x)}}\,dx\right\},
\label{44}\en
and are well defined under our assumptions on $\alpha_j(x)$ and $\beta_j(x)$. Here $N_\varphi$ and $N_\psi$ are normalization constants which will be fixed later. If we now introduce $\varphi_n(x)$ and $\psi_n(x)$ as in (\ref{FB32}),
\be
\varphi_n(x)=\frac{1}{\sqrt{n!}}\,b^n\varphi_0(x),\qquad \psi_n(x)=\frac{1}{\sqrt{n!}}\,{a^\dagger}^n\psi_0(x), \label{45}\en
$n\geq0$, we can prove, \cite{bagJMAA}, the following:

\begin{prop}\label{propstati}
	Calling $\theta(x)=\alpha_a(x)\beta_b(x)+\alpha_b(x)\beta_a(x)$ we have
	\be
	\varphi_n(x)=\frac{1}{\sqrt{n!}}\,\pi_n(x)\varphi_0(x), \qquad \psi_n(x)=\frac{1}{\sqrt{n!}}\,\sigma_n(x)\varphi_0(x),
	\label{46}\en
	$n\geq0$, where $\pi_n(x)$ and $\sigma_n(x)$ are defined recursively as follows:
	\be
	\pi_0(x)=\sigma_0(x)=1,
	\label{47}\en
	and
	\be
	\pi_n(x)=\left(\frac{\theta(x)}{\alpha_a(x)}-\alpha_b'(x)\right)\pi_{n-1}(x)-\alpha_b(x)\pi_{n-1}'(x),
	\label{48}\en
	\be
	\sigma_n(x)=\overline{\left(\frac{\theta(x)}{\alpha_b(x)}-\alpha_a'(x)\right)}\,\sigma_{n-1}(x)-\overline{\alpha_a(x)}\,\sigma_{n-1}'(x),
	\label{49}\en
	$n\geq1$.
\end{prop}

In particular, if $\alpha_a(x)=\alpha_a$ and $\alpha_b(x)=\alpha_b$, both non zero, we have
\be
\pi_n(x)=\sqrt{\left(\frac{\alpha_b}{2\alpha_a}\right)^n}H_n\left(\frac{x+k}{\sqrt{2\alpha_a\alpha_b}}\right), \qquad \sigma_n(x)=\sqrt{\left(\frac{\overline\alpha_b}{2\overline\alpha_a}\right)^n}H_n\left(\frac{x+\overline k}{\sqrt{2\overline\alpha_a\overline\alpha_b}}\right).
\label{411}\en
Here $H_n(x)$ is the $n$-th Hermite polynomial, and the square root of the complex quantities are taken to be their principal determinations. The proof of (\ref{411}) is also contained in \cite{bagJMAA}.

The functions in (\ref{44}), for $\alpha_a(x)=\alpha_a$ and $\alpha_b(x)=\alpha_b$, turn out to be $$\varphi_0(x)=N_\varphi \exp\left\{-\,\frac{1}{\alpha_a}\int\beta_a(x)\,dx\right\}, \qquad \psi_0(x)=N_\psi \exp\left\{-\frac{1}{\overline{\alpha_b}}\int\overline{\beta_b(x)}\,dx\right\},$$ where $\beta_a(x)$ and $\beta_b(x)$, in view of the second equation in (\ref{42}), are only required to satisfy the condition $\alpha_a\beta_b(x)+\alpha_b\beta_a(x)=x+k$. Then we find that $\varphi_{n}(x)\, \overline{\Psi_m(x)}\in\Lc^1(\mathbb{R})$, for all $n,m\geq0$, \cite{bagJMAA}. The proof is based on the fact that $\varphi_{n}(x)\, \overline{\Psi_m(x)}$ is the product of a polynomial of degree $n+m$ times the following exponential
$$
\exp\left\{-\,\int\left(\frac{\beta_a(x)}{\alpha_a}+\frac{\beta_b(x)}{\alpha_b}\right)\,dx\right\}=\exp\left\{-\frac{1}{\alpha_a\alpha_b}\,\int\theta(x)\,dx\right\}=$$
$$=\exp\left\{-\frac{1}{\alpha_a\alpha_b}\,\int(x+k)\,dx\right\}=\exp\left\{-\frac{1}{\alpha_a\alpha_b}\left(\frac{x^2}{2}+kx+\tilde k\right)\right\},
$$
where $\tilde k$ is an integration constant (which is usually fixed to zero). Notice that this is a Gaussian term whenever $\alpha_a\alpha_b>0$. In this case, therefore, it is possible to compute the integral of $\varphi_{n}(x)\, \overline{\Psi_m(x)}$, and this integral is what, with a little abuse of language, we call {\em the scalar product} between $\varphi_{n}(x)$ and $\Psi_m(x)$. We refer to \cite{bagJMAA} for more results and details concerning the biorthogonality (in this extended sense) of $\F_\varphi$ and $\F_\Psi$, both in the case of constant $\alpha_j$, $j=a,b$, and when $\alpha_a(x)$ and $\alpha_b(x)$ are non trivial functions of $x$. Moreover, in \cite{bagJMAA} it is discussed the validity of Assumption $\D$-pbw 3, as well as a possible way to introduce the weak bi-coherent states for the operators in (\ref{41}). What is discussed in \cite{bagJMAA} is relevant, in particular, when $\varphi_{n}(x)$ or $\Psi_m(x)$ are not in $\ltwo$. But, as we will see in the next section, this is not the case here. For this reason we end here our review on WPBs, suggesting the reading of \cite{bagJMAA,bagJPCS2021,bagspringer} for more details, and we move to the explicit model we want discuss in this paper.

\section{The model}\label{sectmodel}

The Hamiltonian we are interested in here is
\be
H=\omega \,b\,a+\lambda(b^2+a^2),
\label{31}\en
in which $a$ and $b$ satisfy (\ref{21}), for some suitable $\D$, dense in $\Hil=\ltwo$, that we will identify later. Here $\omega$ and $\lambda$ are positive real parameters such that $\omega>2\lambda$. As we have discussed in the Introduction, $H$ is a particular version of the Swanson Hamiltonian, \cite{swan}, $H_s=\omega_s c^\dagger c+\alpha c^2+\beta {c^\dagger}^2$, where $\alpha, \beta, \omega_s\in\mathbb{R}$ and $[c,c^\dagger]=\1$, in which the bosonic operators $(c,c^\dagger)$ are replaced by their pseudo-bosonic counterparts, $(a,b)$, and where $\alpha$ coincides with $\beta$. It is clear that both $H_s$ and $H$ are manifestly non self-adjoint, (the latter if $\alpha\neq\beta$). $H$ is not self-adjoint as far as $b^\dagger\neq a$, as will be the case here. The operator $H$ can be diagonalized by means of a simple transformation. Let us introduce a new pair of operators $(A,B)$ as follows:
\be
 A=a\cosh(\theta)+b\sinh(\theta), \qquad B=b\cosh(\theta)+a\sinh(\theta).
\label{32}\en
Then $(A,B)$ are pseudo-bosonic operators, at least formally (at this stage), meaning with this that they also satisfy, as $a$ and $b$, the commutation rule $[A,B]=\1$. We will see later how to make this commutator rigorous, according to our preliminary discussion in Section \ref{sectprel}. Now, if we fix $\theta_0=\frac{1}{2}\tanh^{-1}\left(\frac{2\lambda}{\omega}\right)$, $H$ can be rewritten as
\be
H=\Omega \,B\,A+\gamma\1,
\label{33}
\en
where $\Omega=\frac{\omega}{\cosh(2\theta_0)}$ and $\gamma=-\omega \frac{\sinh^2(\theta_0)}{\cosh(2\theta_0)}$. Now, to be more concrete, we assume that  $a$ and $b$ are {\em shifted} PBs, i.e. that
\be
a=c+\alpha\1,\qquad b=c^\dagger+\beta\1,
\label{34}\en
where $\alpha, \beta\in\mathbb{R}$, $\alpha\neq\beta$, and where $c=\frac{1}{\sqrt{2}}\left(\frac{d}{dx}+x\right)$ and $c^\dagger=\frac{1}{\sqrt{2}}\left(-\frac{d}{dx}+x\right)$ are the usual bosonic operators, densely defined on $\ltwo$. In fact, $D(c)$ and $D(c^\dagger)$ both contain $\scr$, the set of the Schwartz functions. This implies that $a$ and $b$ in (\ref{34})  are densely defined, too. And this is also true for the operators $A$ and $B$ in (\ref{32}), which can be rewritten as follows:
\be
A=\Theta_-\,\frac{d}{dx}+\Theta_+x+\gamma_A\1, \qquad B=-\Theta_-\,\frac{d}{dx}+\Theta_+x+\gamma_B\1,
\label{35}\en
where we have introduced the following quantities:
\be
\Theta_+=\frac{1}{\sqrt{2}}\left(\cosh(\theta_0)+\sinh(\theta_0)\right)=\frac{e^{\theta_0}}{\sqrt{2}}, \qquad \Theta_-=\frac{1}{\sqrt{2}}\left(\cosh(\theta_0)-\sinh(\theta_0)\right)=\frac{e^{-\theta_0}}{\sqrt{2}},
\label{36}\en
and
\be
\gamma_A=\alpha\cosh(\theta_0)+\beta\sinh(\theta_0), \qquad \gamma_B=\beta\cosh(\theta_0)+\alpha\sinh(\theta_0).
\label{37}\en
It is clear then that $A$ and $B$ are of the form in (\ref{41}), with 
\be
\alpha_a(x)=\alpha_b(x)=\Theta_-, \qquad \beta_a(x)=\Theta_+x+\gamma_A, \qquad  \beta_b(x)=\Theta_+x+\gamma_B,
\label{38}\en 
so that the equalities in (\ref{42}) are both satisfied. From (\ref{35}) we have
\be
A^\dagger=-\Theta_-\,\frac{d}{dx}+\Theta_+x+\gamma_A\1, \qquad B^\dagger=\Theta_-\,\frac{d}{dx}+\Theta_+x+\gamma_B\1,
\label{39}\en
since $\Theta_\pm, \gamma_A$ and $\gamma_B$ are all real. Hence
\be
B=A^\dagger+(\gamma_B-\gamma_A)\1.
\label{310}\en
In particular, this last equality shows that $B=A^\dagger$ if and only if $\gamma_A=\gamma_B$, which is surely true if $\alpha=\beta$, see (\ref{37}). But this would imply also that $a=b^\dagger$, which is not interesting for us, since we would go back to ordinary bosonic operators.

The vacua of $A$ and $B^\dagger$ are the following
\be
\varphi_0(x)=N_\varphi\exp\left\{-\frac{\Theta_+}{2\Theta_-}\,x^2-\frac{\gamma_A}{\Theta_-}\,x\right\}, \quad \psi_0(x)=N_\psi\exp\left\{-\frac{\Theta_+}{2\Theta_-}\,x^2-\frac{\gamma_B}{\Theta_-}\,x\right\}, 
\label{311}\en
with $N_\varphi$ and $N_\psi$ normalization constants still to be fixed. Since $\frac{\Theta_+}{2\Theta_-}=\frac{e^{2\theta_0}}{2}$, which is always positive, we conclude that $\varphi_0(x), \psi_0(x)\in\ltwo$. We also observe that $N_\varphi^{-1}\varphi_0(x)$ coincides with $N_\psi^{-1}\psi_0(x)$, by replacing $\gamma_A$ with $\gamma_B$. Using now Proposition \ref{propstati} and (\ref{411}) we deduce that
\be
\varphi_n(x)=\frac{N_\varphi}{\sqrt{2^n\,n!}}\,H_n\left(\frac{x+k}{\sqrt{2}\,\Theta_-}\right)\exp\left\{-\frac{\Theta_+}{2\Theta_-}\,x^2-\frac{\gamma_A}{\Theta_-}\,x\right\},
\label{312}\en
and
\be
\psi_n(x)=\frac{N_\psi}{\sqrt{2^n\,n!}}\,H_n\left(\frac{x+k}{\sqrt{2}\,\Theta_-}\right)\exp\left\{-\frac{\Theta_+}{2\Theta_-}\,x^2-\frac{\gamma_B}{\Theta_-}\,x\right\},
\label{313}\en
where
\be
k=\Theta_-\left(\gamma_A+\gamma_B\right)=\frac{\alpha+\beta}{\sqrt{2}}.
\label{314}\en
Incidentally we observe that the argument of the Hermite polynomials can be rewritten as $\frac{x+k}{\sqrt{2}\,\Theta_-}=e^{\theta_0}(x+k)$, and that, extending what already found for the vacua, $N_\varphi^{-1}\varphi_n(x)$ coincides with $N_\psi^{-1}\psi_n(x)$ replacing $\gamma_A$ with $\gamma_B$, also for $n>0$. It is clear that $\varphi_n(x), \psi_n(x)\in\ltwo$, for all $n\geq 0$ so that, in agreement with what we have seen in Section \ref{sect21}, $\varphi_{n}(x)\, \overline{\Psi_m(x)}\in\Lc^1(\mathbb{R})$, for all $n,m\geq0$. Restricting to real values of $N_\varphi$ and $N_\psi$, and taking
$$
N_\varphi\,N_\psi=\frac{e^{-\frac{k^2}{4\Theta_-}}}{(2\pi)^{1/4}\sqrt{\Theta_-}},
$$
we deduce that the sets $\F_\varphi=\{\varphi_n(x)\}$ and $\F_\psi=\{\psi_n(x)\}$ are biorhonormal:
\be
\langle \varphi_n,\psi_m\rangle=\delta_{n,m}.
\label{315}\en
In the following we will choose
\be
N_\varphi=N_\psi=\frac{e^{-\frac{k^2}{\Theta_-}}}{\sqrt{2\pi}\,\Theta_-}.
\label{316}\en
With this choice, $\varphi_n(x)$ returns $\psi_n(x)$ replacing $\gamma_A$ with $\gamma_B$. The norm of these functions can be easily deduced by adopting to the present case similar computations as those given, for instance, in \cite{baginbagbook}, and which will not be repeated here. In particular we find
\be
\|\varphi_n\|^2=e^{\frac{1}{2}(7\gamma_A^2-\gamma_B^2-2\gamma_A\gamma_B)}\,L_n(-(\gamma_B-\gamma_A)^2),
\label{317}\en
where $L_n$ is a Laguerre polynomial. It is clear that $\|\psi_n\|^2$ can be deduced from (\ref{317}) by replacing $\gamma_A$ with $\gamma_B$.

We see that the argument of $L_n$ is strictly negative, for all $\gamma_A\neq\gamma_B$, so that we can use the following asymptotic (in $n$) formula, \cite{szego},
\be
L_n(x)\simeq\frac{e^x}{2\sqrt{\pi}\,(-x)^{1/4}}\,\frac{e^{2\sqrt{-nx}}}{n^{1/4}},
\label{316bis}\en
which is true if $x<0$. Then, since  $\|\varphi_n\|\|\psi_n\|\rightarrow\infty$, a standard argument shows that $(\F_\varphi,\F_\psi)$ are biorthonormal sets, but none of the two sets is a basis, \cite{davies,baginbagbook}. However, \cite{kolfom}, these two sets are both complete in $\ltwo$. Hence $\Lc_\varphi=l.s.\{\varphi_n\}$ and $\Lc_\psi=l.s.\{\psi_n\}$, the linear spans of the functions $\varphi_n(x)$ and of $\psi_n(x)$, are dense in $\ltwo$.
Moreover, they are $\G$-quasi bases, see (\ref{FB35}), where $\G$ is the following set:
\be
\G=\left\{f(x)\in\ltwo:\,\, e^{qx}f(x)\in\ltwo, \, \forall q\in\mathbb{R}\right\}.
\label{318}\en
This set is dense in $\ltwo$, since it contains $D(\mathbb{R})$, the set of all the compactly supported, $C^\infty$, functions, which is dense in $\ltwo$. To check formula (\ref{FB35}) we first observe that, using (\ref{312}), with the change of variable $y=\frac{x+k}{\sqrt{2}\,\Theta_-}$,
$$
\langle f,\varphi_n\rangle=\int_{\mathbb{R}}\overline{f(x)}\,\varphi_n(x)\,dx= N_\varphi \sqrt{2}\,\pi^{1/4}\Theta_-e^{-\frac{1}{4}(\gamma_A+\gamma_B)(\gamma_B-3\gamma_A)}\langle f_\varphi,e_n\rangle,
$$
where 
$$
f_\varphi(x)=f(\sqrt{2}\,\Theta_--k)\,e^{-x(\gamma_A-\gamma_B)/\sqrt{2}}, \qquad e_n(x)=\frac{1}{\sqrt{2^n\,n! \sqrt{\pi}}} H_n(x)e^{-x^2/2}.
$$
It is well known that $\F_e=\{e_n(x)\}$ is the orthonormal basis of eigenstates of the quantum harmonic oscillator, \cite{mer,mess}. As for $f_\varphi(x)$, this is a square integrable function since $f(x)\in\G$: $f_\varphi(x)\in\ltwo$. With the same change of variable, if $g(x)\in\G$, we can check that
$$
\langle \psi_n,g\rangle=\int_{\mathbb{R}}\overline{\psi_n(x)}\,g(x)\,dx= N_\psi \sqrt{2}\,\pi^{1/4}\Theta_-e^{-\frac{1}{4}(\gamma_A+\gamma_B)(\gamma_A-3\gamma_B)}\langle e_n, g_\psi\rangle,
$$
where
$$
g_\psi(x)=g(\sqrt{2}\,\Theta_--k)\,e^{-x(\gamma_B-\gamma_A)/\sqrt{2}}, 
$$
which is also in $\ltwo$. Now, using the closure relation of the set $\F_e$, we get
$$
\sum_{n=0}^\infty\langle f,\varphi_n\rangle\langle \psi_n,g\rangle=2N_\varphi N_\psi\sqrt{\pi}\,\Theta_-^2\,e^{\frac{1}{2}(\gamma_A+\gamma_B)^2}\,\sum_{n=0}^\infty\langle f_\varphi,e_n\rangle\langle e_n,g_\psi\rangle=
$$
$$
=2N_\varphi N_\psi\sqrt{\pi}\,\Theta_-^2\,e^{\frac{1}{2}(\gamma_A+\gamma_B)^2}\,\langle f_\varphi,g_\psi\rangle.
$$
Next, with the change of variable $y=\sqrt{2}\,\Theta_-\,x-k$, we find that $\langle f_\varphi,g_\psi\rangle=\frac{1}{2\,\Theta_-}\langle f,g\rangle$, which is clearly well defined since $f(x),g(x)\in\ltwo$. Using now (\ref{316}) it is easy to conclude that $\sum_{n=0}^\infty\langle f,\varphi_n\rangle\langle \psi_n,g\rangle=\langle f,g\rangle$. The proof of the other identity in  (\ref{FB35}) is analogous, and will not be repeated. The conclusion is therefore the following: the families $\F_\varphi$ and $\F_\psi$, made of square integrable eigenvectors of, respectively, $N=BA$ and $N^\dagger$, $N\varphi_n=n\varphi_n$ and $N^\dagger\Psi_n=n\Psi_n$, are: (i) complete in $\ltwo$; (ii) biorthonormal; (iii) not bases for $\ltwo$; (iv) $\G$-quasi bases. 

The results deduced in this section allow us to conclude that the pair $(A,B)$ in (\ref{35}) are indeed $\G_0$-pb operators in the sense of Definition \ref{def21}, where $\G_0=\{h(x)\in\G:\, h(x)\in C^\infty\}$. Indeed, $\G_0$ is also dense in $\ltwo$, and, $\forall f(x)\in\G_0$ $[A,B]f(x)=f(x)$. Moreover, $\G_0$  is stable under the action of $A$, $B$, and of their adjoint, and both $A$ and $B^\dagger$ admit vacua in $\G_0$, see (\ref{311}). Finally, Assumption $\D$-pbw 3 is satisfied on (the larger set) $\G$.

\section{Bi-coherent states}\label{sectbcs}

In \cite{bagspringer}, and in some of the references therein, the construction of a special class of coherent states, the so-called bi-coherent states, has been discussed in many details for several classes of pseudo-bosonic operators, and with different techniques. In this section we will consider three such constructions, and compare the respective results.

The first approach we want to consider is based on a theorem, first given in \cite{bagproc}, which can be found in its most recent form in \cite{bagspringer}. We give here this result, without proof.

Let us consider two biorthogonal families of vectors, $\F_{\tilde\varphi}=\{\tilde\varphi_n\in\Hil, \, n\geq0\}$ and $\F_{\tilde\Psi}=\{\tilde\Psi_n\in\Hil, \, n\geq0\}$ which are ${\cal M}$
-quasi bases for some dense subset ${\cal M}$ of $\Hil$,  as in (\ref{FB35}). Consider an increasing sequence of real numbers $\alpha_n$ satisfying the  inequalities $0=\alpha_0<\alpha_1<\alpha_2<\ldots$, and let $\overline\alpha$ be the limit of $\alpha_n$ for $n$ diverging. We further consider two operators, $\tilde A$ and $\tilde B^\dagger$, which act as lowering operators respectively on $\F_{\tilde\varphi}$ and $\F_{\tilde\Psi}$ in the following way:
\be
\tilde A\,\tilde\varphi_n=\alpha_n\tilde\varphi_{n-1}, \qquad \tilde B^\dagger\,\tilde\Psi_n=\alpha_n\tilde\Psi_{n-1},
\label{50}\en
for all $n\geq1$, with $\tilde A\,\tilde\varphi_0=\tilde B^\dagger\,\tilde\Psi_0=0$. These are the lowering equations which replace those in (\ref{FB33}), which can be recovered if $\alpha_n=\sqrt{n}$ and if $\tilde A$ and $\tilde B$ obey (\ref{21}). Then

\begin{thm}\label{theo1}
	Assume that four strictly positive constants $A_\varphi$, $A_\Psi$, $r_\varphi$ and $r_\Psi$ exist, together with two strictly positive sequences $M_n(\varphi)$ and $M_n(\Psi)$, for which
	\be
	\lim_{n\rightarrow\infty}\frac{M_n(\varphi)}{M_{n+1}(\varphi)}=M(\varphi), \qquad \lim_{n\rightarrow\infty}\frac{M_n(\Psi)}{M_{n+1}(\Psi)}=M(\Psi),
	\label{51}\en
	where $M(\varphi)$ and $M(\Psi)$ could be infinity, and such that, for all $n\geq0$,
	\be
	\|\tilde\varphi_n\|\leq A_\varphi\,r_\varphi^n M_n(\varphi), \qquad \|\tilde\Psi_n\|\leq A_\Psi\,r_\Psi^n M_n(\Psi).
	\label{52}\en
	Then, putting $\alpha_0!=1$ and $\alpha_k!=\alpha_1\alpha_2\cdots\alpha_k$, $k\geq1$, the following series:
	\be
	N(|z|)=\left(\sum_{k=0}^\infty\frac{|z|^{2k}}{(\alpha_k!)^2}\right)^{-1/2},
	\label{53}\en
	\be
	\varphi(z)=N(|z|)\sum_{k=0}^\infty\frac{z^k}{\alpha_k!}\tilde\varphi_k,\qquad \Psi(z)=N(|z|)\sum_{k=0}^\infty\frac{z^k}{\alpha_k!}\tilde\Psi_k,
	\label{54}\en
	are all convergent inside the circle $C_\rho(0)$ in $\mathbb{C}$ centred in the origin of the complex plane and of radius $\rho=\overline\alpha\,\min\left(1,\frac{M(\varphi)}{r_\varphi},\frac{M(\Psi)}{r_\Psi}\right)$. Moreover, for all $z\in C_\rho(0)$,
	\be
	\tilde A\varphi(z)=z\varphi(z), \qquad \tilde B^\dagger \Psi(z)=z\Psi(z).
	\label{55}\en
	Suppose further that a measure $d\lambda(r)$ does exist such that
	\be
	\int_0^\rho d\lambda(r) \,r^{2k}=\frac{(\alpha_k!)^2}{2\pi},
	\label{56}\en
	for all $k\geq0$. Then, putting $z=re^{i\theta}$ and calling $d\nu(z,\overline z)=N(r)^{-2}d\lambda(r)d\theta$, we have
	\be
	\int_{C_\rho(0)}\left<f,\Psi(z)\right>\left<\varphi(z),g\right>d\nu(z,\overline z)=
	\int_{C_\rho(0)}\left<f,\varphi(z)\right>\left<\Psi(z),g\right>d\nu(z,\overline z)=
	\left<f,g\right>,
	\label{57}\en
	for all $f,g\in{\cal M}$.
	
\end{thm}

We refer to \cite{bagspringer} for several comments on this theorem. Here we just want to show how to apply this result to our particular operators $A$ and $B$ in (\ref{35}), and to the vectors $\varphi_n(x)$ and $\psi_n(x)$ in (\ref{312}) and (\ref{313}). In this particular situation, of course, $\alpha_n=\sqrt{n}$. 

Using (\ref{316}) and (\ref{316bis}), it is possible to check that
\be
\|\varphi_n\|\simeq c(\gamma_A,\gamma_B)\,\frac{e^{\sqrt{n}|\gamma_A-\gamma_B|}}{n^{1/8}}, \qquad \|\psi_n\|\simeq c(\gamma_B,\gamma_A)\,\frac{e^{\sqrt{n}|\gamma_B-\gamma_A|}}{n^{1/8}},
\label{58}\en
where we have introduced the (inessential) constant 
$$
c(\gamma_A,\gamma_B)=\frac{1}{\sqrt{2\sqrt{\pi|\gamma_B-\gamma_A|}}}\,e^{\frac{1}{2}(3\gamma_A^2-\gamma_B)}.
$$
Therefore, (\ref{52}) are satisfied if we put
$$
A_\varphi=c(\gamma_A,\gamma_B); \qquad A_\psi=c(\gamma_B,\gamma_A); \qquad M_n(\varphi)=M_n(\psi)=\frac{1}{n^{1/8}}, 
$$
and 
$$
r_\varphi=r_\psi=e^{n|\gamma_A-\gamma_B|}.
$$
Hence $M(\varphi)=M(\psi)=1$, and $\rho=\infty$. Then, for our operators $A$ and $B$, the series in (\ref{53}) and (\ref{54}) converge in all of $\mathbb{C}$. Also, in this case the moment problem in (\ref{56}) can be solved, and  $d\nu(z,\overline z)=\frac{1}{\pi}\,r dr\,d\theta$. Since in this case $N(|z|)=e^{-|z|^2/2}$, we write (\ref{54}) as follows
\be
\varphi(z;x)=e^{-|z|^2/2}\,\sum_{l=0}\frac{z^l}{\sqrt{l!}}\,\varphi_l(x), \qquad \psi(z;x)=e^{-|z|^2/2}\,\sum_{l=0}\frac{z^l}{\sqrt{l!}}\,\psi_l(x),
\label{59}\en
where we have put in evidence the role of both $x$ and $z$ in the definition of the states. Theorem \ref{theo1} guarantees that these vectors exist in $\ltwo$, $\forall z\in\mathbb{C}$, produce a resolution of the identity on the set $\G$ in (\ref{318}), see (\ref{57}), and are eigenstates of $A$ and $B^\dagger$, respectively, with eigenvalue $z$, see (\ref{55}). 

It is possible to find a more compact expression for $\varphi(z;x)$ and $\psi(z;x)$. For that we need the well known formula of the generating function for the Hermite polynomials:
\be
\sum_{l=0}^\infty \,\frac{t^l}{l!}\,H_l(x)=\exp\left(-t^2+2tx\right).
\label{510}\en
Now, replacing (\ref{312}) in (\ref{59}),
 we have
 $$
 \varphi(z;x)=N_\varphi e^{-|z|^2/2}\,\exp\left\{-\frac{\Theta_+}{2\Theta_-}\,x^2-\frac{\gamma_A}{\Theta_-}\,x\right\}\sum_{l=0}\frac{z^l}{2^l \,l!}\,\,H_l\left(\frac{x+k}{\sqrt{2}\,\Theta_-}\right),
 $$
 which, using (\ref{316}) and (\ref{510}), produces, after some algebra
 \be
 \varphi(z;x)=\frac{e^{-iz_rz_i+i\frac{z_r}{\Theta_-}(x+k)}}{(2\pi)^{1/4}\sqrt{\Theta_-}}\,\exp\left\{-\frac{k^2}{4\Theta_-^2}-z_r^2-\frac{\Theta_+}{2\Theta_-}\,x^2-\frac{\gamma_A}{\Theta_-}\,x+\frac{z_r}{\Theta_-}(x+k)\right\}.%\,\exp\left\{-iz_rz_i+i\frac{z_r}{\Theta_-}(x+k)\right\}
 \label{511}\en
 Here $z=z_r+iz_i$ and we have separated the phase of $\varphi(z;x)$ from the rest of the function. In a similar way we find
 \be
 \psi(z;x)=\frac{e^{-iz_rz_i+i\frac{z_r}{\Theta_-}(x+k)}}{(2\pi)^{1/4}\sqrt{\Theta_-}}\,\exp\left\{-\frac{k^2}{4\Theta_-^2}-z_r^2-\frac{\Theta_+}{2\Theta_-}\,x^2-\frac{\gamma_B}{\Theta_-}\,x+\frac{z_r}{\Theta_-}(x+k)\right\}.%\,\exp\left\{-iz_rz_i+i\frac{z_r}{\Theta_-}(x+k)\right\},
 \label{512}\en
which coincides with $ \varphi(z;x)$ with the usual exchange\footnote{We remind that $k$ is invariant under this exchange, see (\ref{314}), and so are $\Theta_\pm$, see (\ref{36}).} $\gamma_A\rightleftarrows\gamma_B$.

\vspace{2mm}

We can now check that the same states, apart from the phases, can be found if we look for the solutions of the eigenvalue equations of the type given in (\ref{55}). In particular, if we call $\tilde\varphi(z;x)$ the eigenvalue of the operator $A$ in (\ref{35}), i.e. the solution of $$\frac{d}{dx}\tilde\varphi(z;x)=\frac{1}{\Theta_-}\left(z-\Theta_+\,x-\gamma_A\right),$$
we easily find
\be
\tilde\varphi(z;x)=K_\varphi\,\exp\left\{\frac{1}{\Theta_-}\left((z-\gamma_A)x-\frac{\Theta_+}{2}\,x^2\right)\right\},
\label{513}\en
while the solution of $B^\dagger\tilde\psi(z;x)=z\tilde\psi(z;x)$ is
\be
\tilde\psi(z;x)=K_\psi\,\exp\left\{\frac{1}{\Theta_-}\left((z-\gamma_B)x-\frac{\Theta_+}{2}\,x^2\right)\right\},
\label{514}\en
where $K_\varphi$ and $K_\psi$ are (partly) fixed by the condition $\langle\tilde\varphi,\tilde\psi\rangle=1$. A possible (non unique) solution can be obtained using standard gaussian integration:
\be
K_\varphi=K_\psi=\frac{1}{(2\pi)^{1/4}\,\sqrt{\Theta_-}}\exp\left\{-z_r^2+\frac{z_r\,k}{\Theta_-}-\frac{k^2}{4\Theta_-^2}\right\}.
\label{515}\en
If we now compare $|\varphi(z;x)|$ with $|\tilde\varphi(z;x)|$, they coincide. Analogously, $|\psi(z;x)|=|\tilde\psi(z;x)|$. Then the procedure proposed by Theorem \ref{theo1} is equivalent to solving a simple (first order) differential equation, as it should.

This is not yet the end of the story. Indeed, it is also possible to rewrite our bi-coherent states by making use of certain displacement-like operators. In fact, using the results given in \cite{bagspringer}, which are based on the estimates in (\ref{58}), it is possible to check that the series $\sum_{l=0}^\infty\frac{1}{l!}(\alpha A+\beta B)^l\,\varphi_k(x)$ and $\sum_{l=0}^\infty\frac{1}{l!}(\alpha B^\dagger+\beta A^\dagger)^l\,\psi_k(x)$ are both convergent for all possible complex $\alpha,\beta$ and  $\forall k\geq0$. This means that we can introduce two densely defined operators, $\widetilde V(\alpha,\beta)$ and $\widetilde W(\alpha,\beta)$, as follows:
\be
\widetilde V(\alpha,\beta) f=\sum_{l=0}^\infty\frac{1}{l!}(\alpha A+\beta B)^l\,f, \qquad \widetilde W(\alpha,\beta)g=\sum_{l=0}^\infty\frac{1}{l!}(\alpha B^\dagger+\beta A^\dagger)^l\,g,
\label{516}\en
$\forall f\in\Lc_\varphi$ and $\forall g\in\Lc_\psi$. For obvious reasons, it is natural to write
$$
\widetilde V(\alpha,\beta) f=e^{\alpha A+\beta B}\,f, \qquad  \widetilde W(\alpha,\beta)g=e^{\alpha B^\dagger+\beta A^\dagger}g,
$$
for the same $f$ and $g$ as above. Now, see \cite{bagspringer}, our bi-coherent states above can be rewritten in terms of these operators. In particular,
\be
\varphi(z;x)=\widetilde V(-\overline{z},z)\varphi_0(x), \qquad \psi(z;x)=\widetilde W(-\overline{z},z)\psi_0(x),
\label{517}\en
which is still a third way to express the bi-coherent states for our extended Swanson model. In other words, $\widetilde V(-\overline{z},z)$ and $\widetilde W(-\overline{z},z)$ play here the role of the unitary displacement operator for ordinary coherent states.

\begin{figure}[ht]
	\begin{center}
		\includegraphics[width=0.48\textwidth]{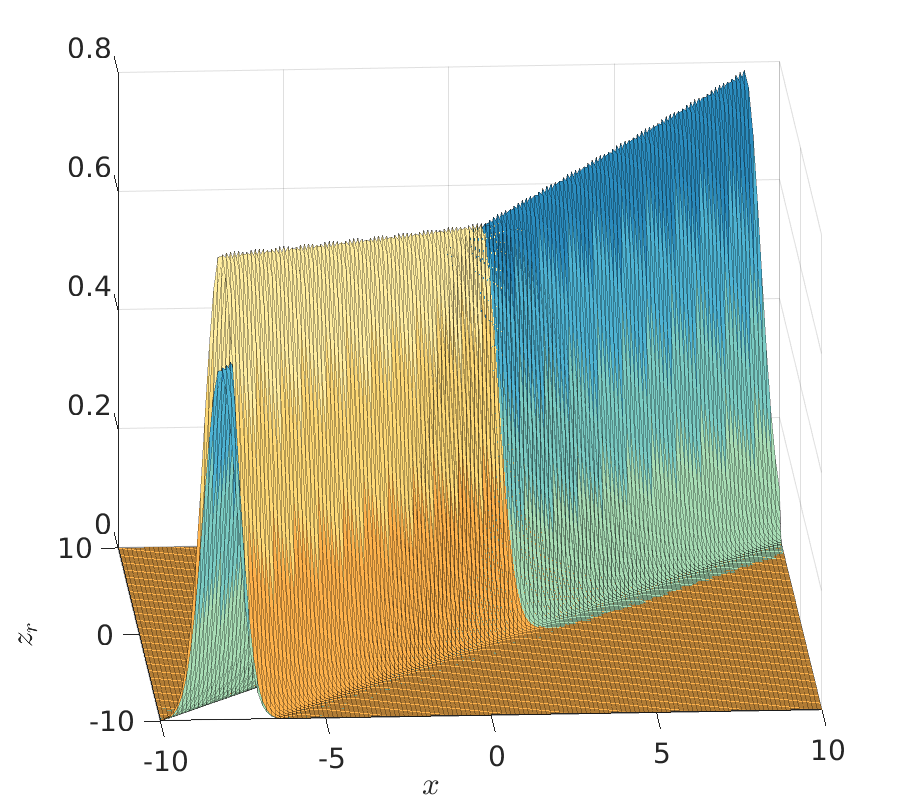}\hspace{1mm} %
		\includegraphics[width=0.48\textwidth]{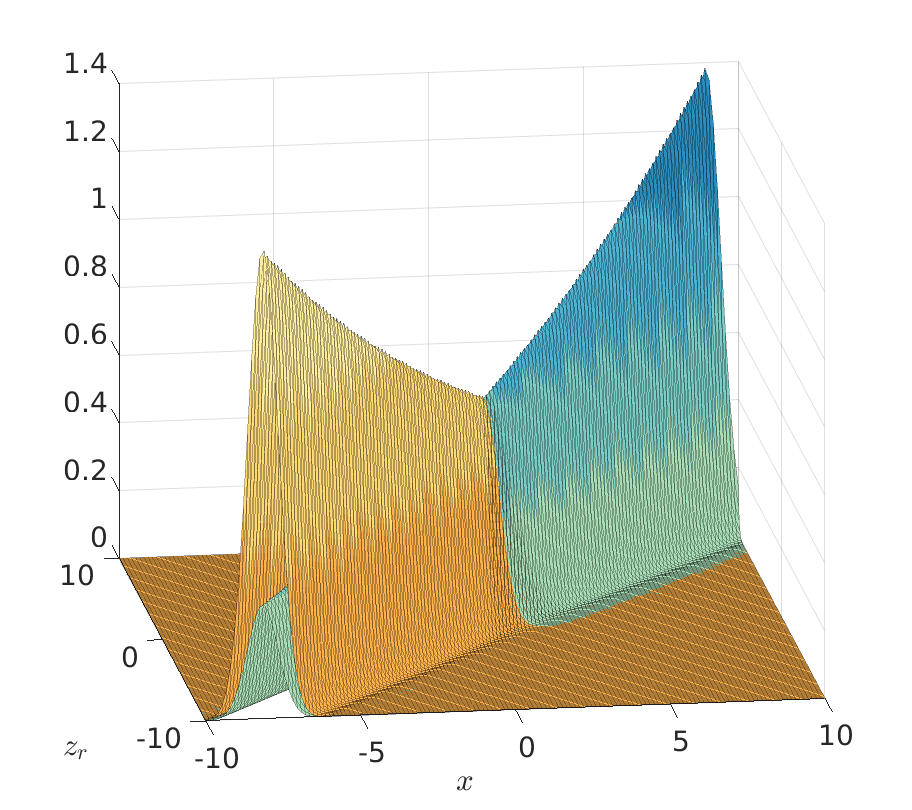}\hfill\\[0pt]
		\includegraphics[width=0.48\textwidth]{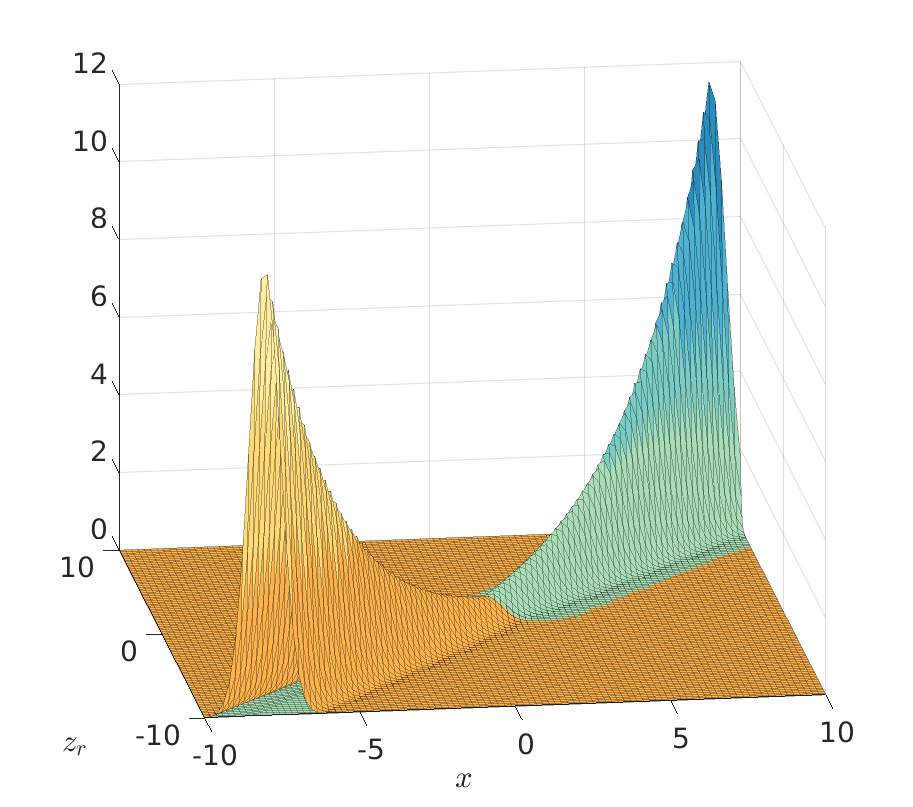}\hspace{1mm} %
		\includegraphics[width=0.48\textwidth]{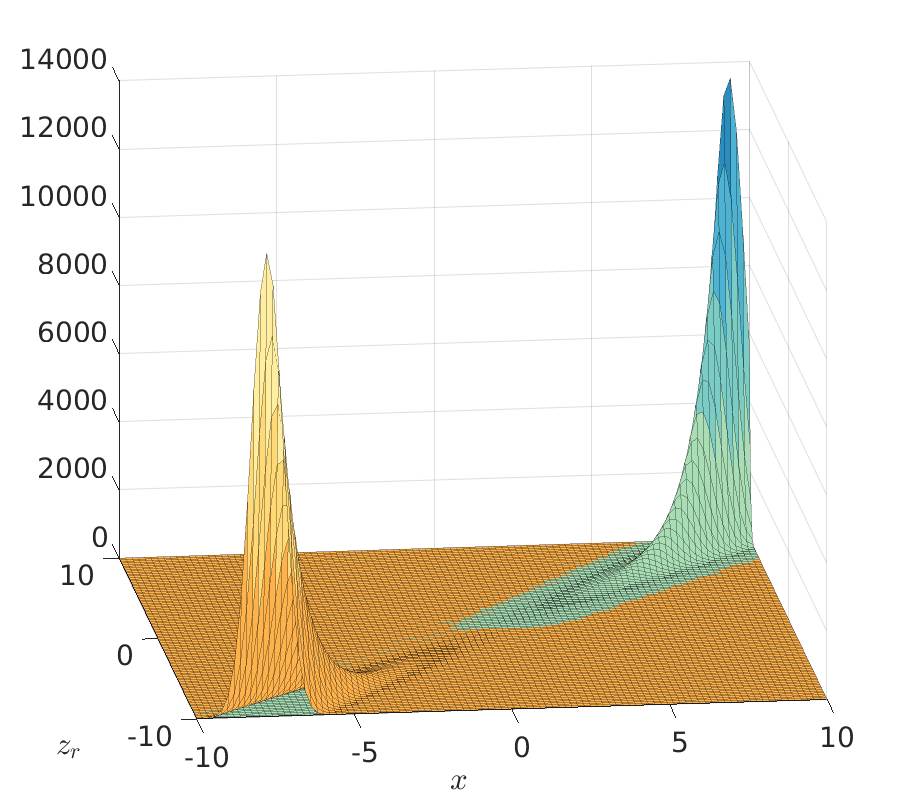}
	\end{center}
	\caption{{\protect\footnotesize $|\psi(z;x)|^2$ (orangish) and $|\varphi(z;x)|^2$ (blueish) in (\ref{511}) and (\ref{512}) for different $\alpha$ and $\beta$ and for $\lambda=0.1$ and $\omega=0.5$: (top left) $\alpha=0.3$, $\beta=0.31$; (top right) $\alpha=0.3$, $\beta=0.35$; (bottom left) $\alpha=0.3$, $\beta=0.5$; (bottom right) $\alpha=0.3$, $\beta=1$. }}
	\label{fig1}
\end{figure}

We plot in Figure \ref{fig1} the square moduli of $\psi(z;x)$ and $\varphi(z;x)$ for $\lambda=0.1$ and $\omega=0.5$ and for different choices of $\alpha$ and $\beta$. We observe that our choice of $\omega$ and $\lambda$ satisfies the constraint given at the beginning of Section \ref{sectmodel}, $\omega>2\lambda$. We further observe that the different choices of $\alpha$ and $\beta$ considered in the figure correspond, see (\ref{34}), to operators $a$ and $b^\dagger$ which are more and more different. This increasing difference is reflected in the plots of the bi-coherent states, which tend to move away more and more one from the other when $\beta-\alpha$ increases. This is essentially the same behaviour we have already observed in several other concrete examples of bi-coherent states, see \cite{bagspringer}.  

%\begin{figure}[ht]
%	\begin{center}
%		\includegraphics[width=0.48\textwidth]{figures/bcs_al03_be031la01ome05.pdf}\hspace{1mm} %
%		\includegraphics[width=0.48\textwidth]{figures/bcs_al03_be035la01ome05.pdf}\hfill\\[0pt]
%		\includegraphics[width=0.48\textwidth]{figures/bcs_al03_be05la01ome05.pdf}\hspace{1mm} %
%		\includegraphics[width=0.48\textwidth]{figures/bcs_al03_be1la01ome05.pdf}
%	\end{center}
%	\caption{{\protect\footnotesize $|\psi(z;x)|^2$ (in orange) and $|\varphi(z;x)|^2$ (in blue) in (\ref{511}) and (\ref{512}) for different $\alpha$ and $\beta$ and for $\lambda=0.1$ and $\omega=0.5$: (top left) $\alpha=0.3$, $\beta=0.31$; (top right) $\alpha=0.3$, $\beta=0.35$; (bottom left) $\alpha=0.3$, $\beta=0.5$; (bottom right) $\alpha=0.3$, $\beta=1$. }}
%	\label{fig1}
%\end{figure}
%

%\section{Path integrals}\label{sectpath}

\section{Conclusions}\label{sectconcl}

In this paper we have discussed how a particular fully pseudo-bosonic Swanson model can be introduced and how its Hamiltonian $H$ can be diagonalized. We have also found the eigensystem of $H^\dagger$, using the general framework and results deduced in the context of PBs. We have constructed, using different approaches, the bi-coherent states for the model, we have compared the results and we have deduced some of their properties.

Several extensions of the model proposed here could be considered: first of all, rather than the Hamiltonian $H$ in (\ref{31}) one could consider the more general, but still quadratic, operator $\hat H=\omega \,b\,a+\lambda_bb^2+\lambda_aa^2$. Further, one could analyze the role, if any, of the WPBs, trying to see how much of the results given here can be extended to a distributional settings, as proposed in \cite{bagJPA2020,bagJMAA}. And, last but not least, it would be interesting to apply the results deduced here in the computation of some propagator, by making use of the properties of bi-coherent states of the model, in the same line as in \cite{bagfein,bagspringer}. These are part of our programs for the future.

\section*{Acknowledgements}

The author acknowledges partial financial support from Palermo University (via FFR2021 "Bagarello") and from G.N.F.M. of the INdAM.

%
%
%\renewcommand{\theequation}{A.\arabic{equation}}
%
%\section*{Appendix A: proof of the biorthonormality}\label{appendixA}
%
%
%cxvcxv
%
% 
%
%
%
%\renewcommand{\theequation}{B.\arabic{equation}}
%
%\section*{Appendix B: proof of Lemma \ref{lemma1}}\label{appendixB}
%
%
%dsfdsfs

\end{document}